\documentclass[a4paper,11pt]{article}

\usepackage{pos}
\usepackage[capitalise]{cleveref}
\usepackage[utf8]{inputenc}
\usepackage{graphicx}
\usepackage{subcaption}
\usepackage{hyperref}

\graphicspath{{./figures/}}

\newcommand{\ii}{\mathrm{i}}
\newcommand{\obs}{\mathcal{O}}
\newcommand{\x}{\phantom{-}}
\newcommand{\y}{\phantom{0}}
\newcommand{\z}{\hspace{-0.15cm}}

\title{Kernels and integration cycles in complex Langevin simulations}

\author*[a]{Michael Mandl}
\author[a]{Michael W. Hansen}
\author[b]{Erhard Seiler}
\author[a]{D{\'e}nes Sexty}

\affiliation[a]{Institute of Physics, NAWI Graz, University of Graz,\\ Universitätsplatz 5, 8010 Graz, Austria}
\affiliation[b]{Max-Planck-Institut für Physik (Werner-Heisenberg-Institut), \\ Boltzmannstraße 8,
85748 Garching bei München, Germany}

\emailAdd{michael.mandl@uni-graz.at}
\emailAdd{michael.hansen@uni-graz.at}
\emailAdd{ehs@mpp.mpg.de}
\emailAdd{denes.sexty@uni-graz.at}

\abstract{
	The method of complex Langevin simulations is a tool that can be used to tackle the complex-action problem encountered, for 
	instance, in finite-density lattice quantum chromodynamics or real-time lattice field theories. The method is based 
	on a stochastic evolution of the dynamical degrees of freedom via (complex) Langevin equations, which, however, 
	sometimes converge to the wrong equilibrium distributions. While the convergence properties of the evolution can to 
	some extent be assessed by studying so-called boundary terms, we demonstrate in this contribution that boundary 
	terms on their own are not sufficient as a correctness criterion. Indeed, in their absence complex Langevin 
	simulation results might still be spoiled by unwanted so-called integration cycles. In particular, 
	we elaborate on how the introduction of a kernel into the complex Langevin equation	can -- in principle -- be used to 
	control which integration cycles are sampled in a simulation such that correct convergence is restored.
	}

\FullConference{The 41st International Symposium on Lattice Field Theory (LATTICE2024)\\
 28 July - 3 August 2024\\
Liverpool, UK\\}

\begin{document}
\maketitle

\section{Introduction}\label{sec:introduction}
	The study of quantum chromodynamics (QCD) at non-zero chemical potential is notoriously difficult due to the 
infamous sign or complex-action problem preventing the use of straightforward lattice quantum field
theory methods based on importance sampling. Nonetheless, a proper understanding of the finite-density
region of the QCD phase diagram is of fundamental importance for the correct description of, e.g., 
neutron stars or the heavy-ion collision experiments at RHIC or the LHC.
One particular method that has seen progress in recent years in the context of finite-density QCD
\cite{Sex14,ASS14,FKS15,HS24} as well as various other theories plagued by a sign problem, such as the 
real-time evolution of lattice field theories \cite{BHM23,ARS24}, is the complex Langevin method, which is based on 
a complexified version of the well-known Langevin equation \cite{Par83,Kla83}. It avoids the sign problem by 
trading the complex weight $e^{-S}$ in the conventional Euclidean path integral for a genuine probability density 
in complexified field space. However, the complex Langevin approach does not come without its own set of problems, 
as it, for instance, sometimes produces wrong results despite apparently properly converging to equilibrium. Thus, 
an important milestone in the development of the complex Langevin method would be to establish a sensible criterion 
of correctness that could detect such incorrect convergence. One candidate for a correctness criterion of this kind 
is the study of boundary terms, the appearance of which spoils the formal proof of convergence of the 
method \cite{AJS11,SSS19}. As we demonstrate in this contribution, however, this criterion is not always reliable 
as it may sometimes falsely indicate correct results. The origin of this failure is contributions from unwanted 
so-called integration cycles. In this contribution, we show how to -- in principle -- remove these contributions via the 
introduction of a kernel into the complex Langevin equation.

\section{Complex Langevin equation and boundary terms}\label{sec:cle}
	The (complex) Langevin equation is a stochastic differential equation describing the evolution of a 
physical system in an auxiliary time direction $\tau$, the so-called Langevin time. For a single (complex) 
degree of freedom $z=x+\ii y$, it reads
\begin{equation}\label{eq:cle}
	\frac{d z(\tau)}{d\tau} = 
		- K\frac{\partial S(z)}{\partial z} + \sqrt{K}\eta(\tau)\;,
\end{equation}
where $S$ denotes the action of the theory, $S(x)$, analytically continued to complex arguments and $\eta$ is real 
Gaussian random noise with the properties $\langle\eta(\tau)\rangle=0$ and 
$\langle\eta(\tau)\eta(\tau')\rangle=2\delta(\tau-\tau')$. Moreover, $K$ denotes the so-called kernel that can be 
used to control the convergence properties of the evolution. It is chosen constant, i.e., independent of $z$, here 
for simplicity albeit this choice could easily be relaxed. 

The stochastic evolution \eqref{eq:cle} gives rise to a probability density $P(x,y;\tau)$ in the complex plane
and the complex Langevin approach solves the sign problem if and only if this probability density is equivalent to 
the target complex weight $e^{-S}$ of the original ("path") integral, $Z = \int dx e^{-S(x)}$, in the sense that
\begin{equation}\label{eq:observables}
	\lim_{\tau\to\infty}\int dxdyP(x,y;\tau)\obs(x+\ii y) = 
		\frac{1}{Z}\int dx e^{-S(x)}\obs(x)
\end{equation}
for all holomorphic observables $\obs$. In the case of a real action $S$, a similar convergence condition 
can indeed be proven under very mild assumptions \cite{PW81}. For complex $S$, on the other hand, a sound 
mathematical formulation does not exist and, in fact, the complex Langevin equation in some cases fails to produce 
correct results entirely, as we shall demonstrate in the remainder of this work.

Part of this failure can be traced back to an insufficient decay of $P(x,y;\tau)$ in the complex plane,
which spoils the formal proof of correctness of the complex Langevin approach (relying on integration by parts)
due to the appearance of boundary terms at infinity \cite{AJS11,SSS19}. It is possible in principle to measure 
these boundary terms in a simulation \cite{SSS20} via the observable
\begin{equation}\label{eq:boundary_terms}
	\mathcal{B}_{\obs(z)}(Y) = \left\langle\Theta\left(Y-\vert z\vert\right)L\obs(z)\right\rangle\;,
\end{equation}
where 
$L = \left(\frac{\partial}{\partial z}-\frac{\partial S(z)}{\partial z}\right)K\frac{\partial}{\partial z}$ and 
$Y$ denotes a cutoff one imposes on the distribution in the complex plane in order to alleviate signal-to-noise 
problems. In practice, one looks for a plateau in $\mathcal{B}_{\obs(z)}(Y)$ and extrapolates to
$Y\to\infty$. Indeed, if such a plateau is assumed at a non-vanishing value of $\mathcal{B}_{\obs(z)}$,
one concludes that correct convergence cannot be proven as the integration by parts cannot be performed 
and, thus, that the obtained simulation results must be incorrect.
The converse, however, is not necessarily true. Indeed, as has been shown in \cite{SS19} and as we 
shall substantiate in this contribution, the absence of boundary terms does not automatically imply the 
correct convergence of a complex Langevin simulation. The reason for this is outlined in the subsequent section.
	
\section{Integration cycles}\label{sec:cycles}
	It was shown in \cite{SS19} that vanishing boundary terms do not guarantee the convergence of results 
obtained in a complex Langevin simulation, $\langle\obs\rangle_\mathrm{CL}$, to the correct values
$\langle\obs\rangle_\mathrm{exact}$, but rather, as was surmised already in \cite{Sal93}, that the former are 
given by a linear combination of observables computed along so-called integration cycles $\gamma_i$,
\begin{equation}\label{eq:salcedo_seiler}
	\langle\obs\rangle_\mathrm{CL} = \sum_{i=1}^{N_\gamma}a_i\langle\obs\rangle_{\gamma_i}\;, \quad 
	\sum_{i=1}^{N_\gamma}a_i=1\;,
\end{equation}
where the $a_i$ are complex coefficients independent of the observables and we have defined
\begin{equation}\label{eq:cycle_integrals}
	\langle\obs\rangle_{\gamma_i} := 
		\frac{\int_{\gamma_i}dz e^{-S(z)}\obs(z)}{\int_{\gamma_i}dz e^{-S(z)}}\;.
\end{equation}
The cycles $\gamma_i$ are defined to be integration paths in the complex plane that either connect two 
distinct zeros of the weight $e^{-S}$ (including those at complex infinity) or are closed incontractible 
loops, while $N_\gamma$ denotes the number of such integration cycles that are linearly independent of one 
another. If one defines $\gamma_1$ such that $\langle\obs\rangle_{\gamma_1}=\langle\obs\rangle_\mathrm{exact}$, 
the complex Langevin evolution 
produces correct results if and only if $a_i=\delta_{i,1}$. Since -- in general -- neither the $a_i$ nor the
$\langle\obs\rangle_{\gamma_i}$ are known \emph{a priori}, a natural question that arises is whether one has 
any sort of control over the coefficients in practice in order to meet this condition.

It was found in \cite{OOS89} that the equilibrium solutions produced by the complex Langevin equation can be 
influenced by the choice of kernel $K$ in \eqref{eq:cle}, an observation that was further elaborated on in 
\cite{Sal93}. Indeed, as we show below, the coefficients $a_i$ in \eqref{eq:salcedo_seiler} 
depend on $K$ in some way. To see how, we follow \cite{OOS89} by considering the following simple toy 
model:
\begin{equation}\label{eq:1d}
 S = \frac{\lambda_l}{4}z^4\;, \quad 
 \lambda_l := e^{\ii\pi l/6}\;, \quad l\in\{-5,\,\dots,6\}\;.
\end{equation}
It was shown in \cite{OOS89} that with a trivial kernel, $K=1$, the complex Langevin evolution converges to 
the analytical solution for the expectation value of the observable $\obs(z)=z^2$,
\begin{equation}\label{eq:exact}
	\langle z^2\rangle_{\mathrm{exact}} = 
		\left(\frac{4}{\lambda_l}\right)^{1/2}\frac{\Gamma(3/4)}{\Gamma(1/4)}\;,
\end{equation}
to a satisfactory degree as long as $-2\lesssim l\lesssim2$ but fails outside that interval. We have 
reproduced these results using an improved discretization of the Langevin equation including an adaptive 
step-size algorithm (the precise technical details will be published elsewhere) and we summarize our 
findings, including the corresponding boundary terms $\mathcal{B}_{z^2}$ from \eqref{eq:boundary_terms}, in 
\cref{tab:boundary_terms}.

\begin{table}[t]
	\centering
	\caption{Expectation value $\langle z^2\rangle$ in the model \eqref{eq:1d} for different values of $l$ 
			 (first column), both computed exactly via \eqref{eq:exact} (second column) 
			 and in a complex Langevin simulation (third column). We also show 
			 the boundary term $\mathcal{B}_{z_2}$ for $\langle z^2\rangle$ evaluated at some representative value 
			 $Y_\mathrm{plat}$ where $\mathcal{B}_{z^2}(Y)$ forms a plateau (last column). The numerical 
			 results are rounded to the first significant digit of their estimated jackknife error.}
	\label{tab:boundary_terms}
	\begin{tabular}{|c|c|c|c|}
		\hline
		$l$ & $\langle z^2\rangle_{\mathrm{exact}}$ & $\langle z^2\rangle_{\mathrm{CL}}$ & $\mathcal{B}_{z^2}(Y_\mathrm{plat})$\\
		\hline
		$-5$ & $0.174956+0.652945\,\ii$ & $\z-0.652933(8)+0.17497(2)\,\ii$ & $\z\y$$0.0000(2)-0.0002(3)\,\ii\z$ \\
		$-4$ & $0.337989+0.585414\,\ii$ & $\z\y$$-0.40380(2)+0.44282(2)\,\ii$ & $\z\y\x$$1.37(2)-2.37(2)\,\ii$$\x\z$ \\
		$-3$ & $0.477989+0.477989\,\ii$ & $\z\x\y$$0.41415(2)+0.50696(1)\,\ii$ & $\z\x\x\x$$0.001(2)-0.326(2)\,\ii$$\y\y\y\z$ \\
		$-2$ & $0.585414+0.337989\,\ii$ & $\z\x\y$$0.58543(1)+0.33801(1)\,\ii$ & $\z\y$$-0.00013(6)-0.00006(2)\,\ii$$\x\z$ \\
		$-1$ & $0.652945+0.174956\,\ii$ & $\z\x\x\y$$0.65294(1)+0.174953(5)\,\ii$ & $\z\x$$0.00002(6)+0.00001(1)\,\ii\z$ \\
		$0$ & $0.675978+0.000000\,\ii$ & $\z0.67598(2)+0.00000\,\ii$ & $\z\y$$0.00002(5)+0.00000\,\ii$$\x\x\z$ \\
		$1$ & $0.652945-0.174956\,\ii$ & $\z\x\x\y$$0.65296(2)-0.174962(8)\,\ii$ & $\z\y$$-0.00008(6)+0.00001(1)\,\ii$$\x\z$ \\
		$2$ & $0.585414-0.337989\,\ii$ & $\z\x\y$$0.58542(1)-0.33801(1)\,\ii$ & $\z\y$$-0.00010(6)+0.00008(3)\,\ii$$\x\z$ \\
		$3$ & $0.477989-0.477989\,\ii$ & $\z\x\y$$0.41414(2)-0.50696(1)\,\ii$ & $\z-0.001(1)+0.328(1)\,\ii\z$ \\
		$4$ & $0.337989-0.585414\,\ii$ & $\z\y$$-0.40392(4)-0.44275(9)\,\ii$ & $\z\y$$1.370(5)+2.377(5)\,\ii\z$ \\
		$5$ & $0.174956-0.652945\,\ii$ & $\z-0.652953(9)-0.17495(2)\,\ii$ & $\z\y$$0.0003(1)+0.0001(1)\,\ii\z$ \\
		$6$ & $0.000000+0.675978\,\ii$ & $\z-0.675981(6)-0.00003(3)\,\ii$ & $\z\y$$0.00006(5)+0.00006(7)\,\ii\z$ \\
		\hline
	\end{tabular}
\end{table}

The curious point to note is the following: While the study of boundary terms can detect the success or 
failure of a simulation correctly for most values of $\lambda_l$, it fails for $l=-5$, $5$ and $6$. For 
these values, the boundary terms are very small and most likely consistent with zero within discrete-step-size and 
round-off errors even though the stochastic evolution converges to an equilibrium distribution that gives incorrect results for holomorphic observables. This observation shows that the study of boundary terms is not sufficient in certain cases.
Recall, however, that up to now we have only considered a single observable, $\langle z^2\rangle$, and the 
boundary term associated with it. 

\begin{figure}[t]
	\centering
	\begin{subfigure}{0.495\linewidth}
		\includegraphics[scale=0.45]{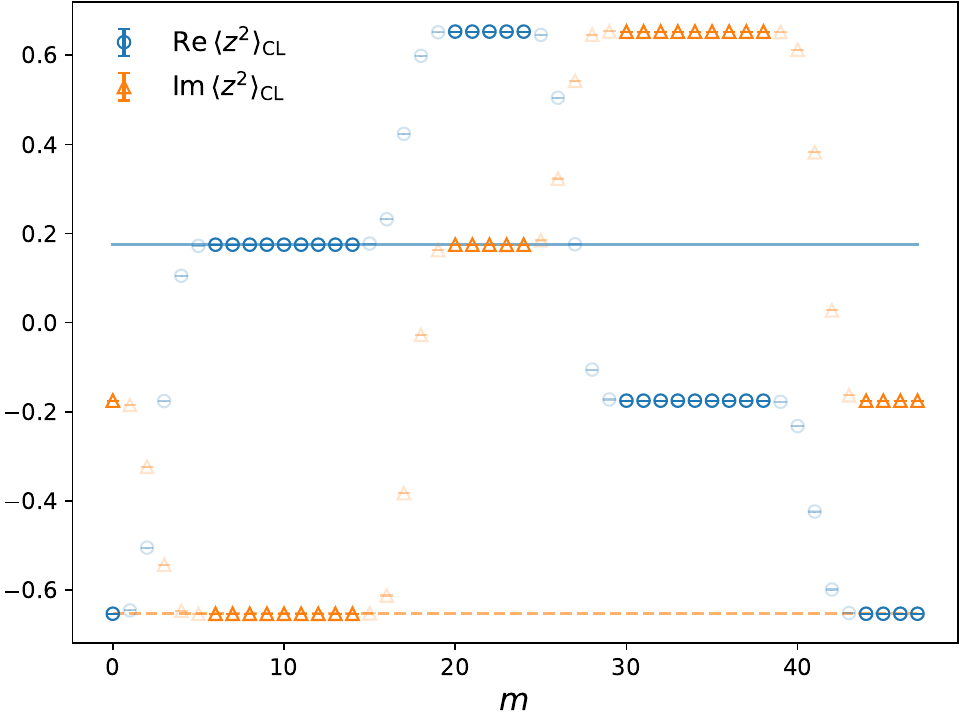}
	\end{subfigure}
	\begin{subfigure}{.495\linewidth}
		\includegraphics[scale=0.45]{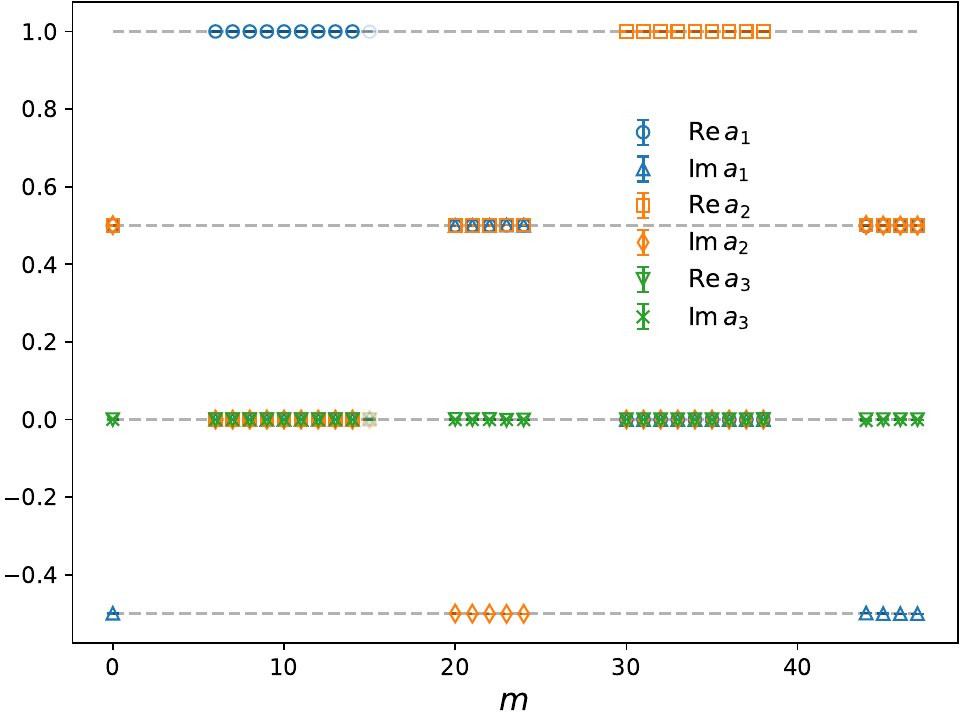}
	\end{subfigure}
	\caption{\emph{Left:} Expectation value $\langle z^2\rangle$ in the model \eqref{eq:1d} for $l=5$ as a 
			 function of $m$ in \eqref{eq:kernel}. We also show the real and imaginary parts of the exact result 
			 \eqref{eq:exact} as solid and dashed horizontal lines, respectively. \emph{Right:} Coefficients $a_i$ 
			 for the same setup, computed from \eqref{eq:salcedo_seiler} using $\langle z\rangle_\mathrm{CL}$, 
			 $\langle z^2\rangle_\mathrm{CL}$, 
			 and $\langle z^4\rangle_\mathrm{CL}$. We only show data points for which the coefficients could be 
			 determined consistently. The dashed horizontal lines mark the values $-0.5$, $0$, $0.5$, and $1$ for 
			 convenience. In both figures, the data points for which the boundary terms are not clearly nonzero 
			 are highlighted.}
	\label{fig:1D}
\end{figure}

Let us now discuss the role of the kernel in \eqref{eq:cle}. By parametrizing it as
\begin{equation}\label{eq:kernel}
	K = K_m := e^{-\ii\pi m/24}\;, \quad m\in\{0,\,\dots,47\}\;,
\end{equation}
we may study the dependence of observables on $m$. Considering $l=5$ in \eqref{eq:1d}, we show the dependence 
of $\langle z^2\rangle$ on $m$ in \cref{fig:1D} (left). Indeed, as was also shown in \cite{OOS89}, there is a plateau 
around $m=10$ on which the analytical result \eqref{eq:exact} is reproduced.\footnote{We mention in passing 
that close to $m=10$ the kernel has the effect of aligning the distribution of $z$ in the complex plane 
with the relevant Lefschetz thimble. For a discussion on the role of Lefschetz thimbles in the context of 
complex Langevin simulations, see, e.g., \cite{Aar13}.} Moreover, we find that on this plateau the boundary terms 
$\mathcal{B}_{z^2}$ are consistent with zero. Notice, however, that there are additional plateaus, around $m=22$, $34$, 
and $46$, respectively, on which boundary terms vanish but the complex Langevin result nonetheless disagrees 
with the exact one. From the theorem \eqref{eq:salcedo_seiler}, we thus conclude that close to $m=10$ one must 
have $a_i=\delta_{i,1}$, as desired, whereas outside this region other integration cycles also contribute 
non-negligibly. Let us now quantify this statement.

To this end, we first count the number of independent integration cycles $N_\gamma$ of the theory 
\eqref{eq:1d}, setting $\lambda_l=1$ for now without loss of generality. It is not hard to see that $e^{-S}$ 
vanishes in four disconnected regions around $z=\pm\infty$ 
and $z=\pm\ii\infty$, respectively. This leaves us (up to homology) with six possible integration cycles in 
total, only $N_\gamma=3$ of which are linearly independent, as can be shown using basic sum rules of line 
integrals. We may, for instance, choose the following basis set of integration cycles:
\begin{equation}\label{eq:basis_cycles}
	\int_{\gamma_1}dz := \int_{-\infty}^\infty dz\;, \quad 
	\int_{\gamma_2}dz := \int_{-\ii\infty}^{\ii\infty} dz\;, \quad
	\int_{\gamma_3}dz := \int_{\ii\infty}^\infty dz\;.
\end{equation}
We emphasize that the precise shape of the integration contours in \eqref{eq:basis_cycles} is irrelevant due to Cauchy's 
theorem as we only consider holomorphic observables; the cycles are thus uniquely specified by their respective 
end points. 

For general $\lambda_l$ in the model \eqref{eq:1d}, we define $\gamma_1$ to be the real line rotated by 
an angle $\theta=\arg(\sqrt[4]{\lambda^{-1}})$. On such an integration contour $S$ is real and positive. The 
other cycles $\gamma_2$ and $\gamma_3$ are then defined analogously via a rotation of the paths given in \eqref{eq:basis_cycles}. The expectation values $\langle z^n\rangle_{\gamma_i}$, defined in 
\eqref{eq:cycle_integrals}, can be computed exactly for all such $\gamma_i$ and positive integers $n$.
This allows 
one to extract the coefficients $a_i$ from \eqref{eq:salcedo_seiler} by measuring a sufficiently large set of 
monomial observables $\langle z^n\rangle_\mathrm{CL}$. Since $\sum_{i=1}^{N_\gamma}a_i=1$ eliminates one (complex) 
degree of freedom, one requires a minimum of $N_\gamma-1=2$ observables for the fit. Note, however, that
$\langle z^{4n+3}\rangle_{\gamma_i}=0$ for all $i$, which excludes the use of such powers. 
Here, we determine the coefficients by measuring $\langle z\rangle_\mathrm{CL}$, $\langle z^2\rangle_\mathrm{CL}$, 
and $\langle z^4\rangle_\mathrm{CL}$ in order to ensure that the fit is stable, while at the same time 
avoiding the large fluctuations encountered for higher powers. Doing so for different values of $m$ at $l=5$, we 
obtain the dependence of $a_i$ on the kernel shown in \cref{fig:1D} (right). 
As expected, we find that around $m=10$ all coefficients except $a_1=1$ vanish. On the contrary, close to 
$m=34$ it is the second cycle that dominates. For other values of $m$ with 
vanishing boundary terms, one finds different linear combinations of the $\langle\obs\rangle_{\gamma_i}$ to 
contribute, all in accordance with \eqref{eq:salcedo_seiler}. Notice that even for certain kernels where there are 
non-zero boundary terms, in which case \eqref{eq:salcedo_seiler} has not been proven to apply, we 
nonetheless obtain more or less reasonable fits. It is also striking to observe that $a_3$ is consistent with zero 
for all values of $m$ shown. This fact is noteworthy since it occurs only when averaging over multiple 
independent simulation runs. On the contrary, for appropriate subsets of runs with similar initial configurations, we 
observe non-ergodicity in certain cases, which may then give rise to non-vanishing values of $a_3$. We
shall discuss this issue in more detail in a forthcoming publication.

These observations make clear that a kernel $K$ in \eqref{eq:cle} can have a non-trivial influence on the 
integration cycles that are sampled in a complex Langevin simulation. In the simple example considered here, it is 
not hard to understand why a kernel of the form \eqref{eq:kernel} with $m\approx10$ gives rise to correct results 
in the theory \eqref{eq:1d} with $l=5$. After all, with this choice the complex Langevin dynamics \eqref{eq:cle} 
becomes equivalent to real Langevin dynamics (albeit on a rotated "real" axis as defined above) \cite{OOS89}, which can be shown to 
converge. In more realistic theories, however, one can only guess what an appropriate kernel could be. Hence, a 
better understanding of how (if at all) kernels can influence the convergence, as well as which integration cycles 
are sampled in a simulation, would be very much desirable. Obviously, in more complicated theories one cannot 
simply compute the $\langle\obs\rangle_{\gamma_i}$ and then fit the coefficients $a_i$, as the former would 
already amount to solving the theory altogether. In fact, in general it is very difficult to determine what the independent 
cycles $\gamma_i$ are in the first place. Moreover, the theorem \eqref{eq:salcedo_seiler} was only proven for a single degree of 
freedom to begin with. As a first step towards a better understanding of the role of integration cycles in 
realistic theories, we thus devote the remainder of this work to a numerical study of the validity of 
\eqref{eq:salcedo_seiler} for two degrees of freedom.
	
\section{Integration cycles in two dimensions}\label{sec:cycles_higher_dimensions}
	Let us consider the following straightforward extension of the model \eqref{eq:1d} to two degrees of freedom,
$z_1$ and $z_2$:
\begin{equation}\label{eq:2d}
	S = \frac{\lambda_l}{4}(z_1^2+z_2^2)^2\;,
\end{equation}
where we once more parametrize $\lambda_l$ as in \eqref{eq:1d}. The concept of integration cycles is
generalized to $d$ dimensions rather straightforwardly as $d$-dimensional integration paths in $\mathbb{C}^d$ that 
either connect two distinct zeros of $e^{-S}$ or are closed and incontractible. As before, in the model 
\eqref{eq:2d} we do not need to worry about the latter, nor about finite zeros, as $e^{-S}$ has no singularities 
and only vanishes as $\vert z_i\vert\to\infty$. In particular, there are zeros whenever $\mathrm{Re}\,S\to\infty$ 
as $\vert z_i\vert\to\infty$. To find the independent integration cycles of \eqref{eq:2d}, we shall first 
determine these zeros.

\begin{figure}[t]
	\centering
	\includegraphics[scale=0.47]{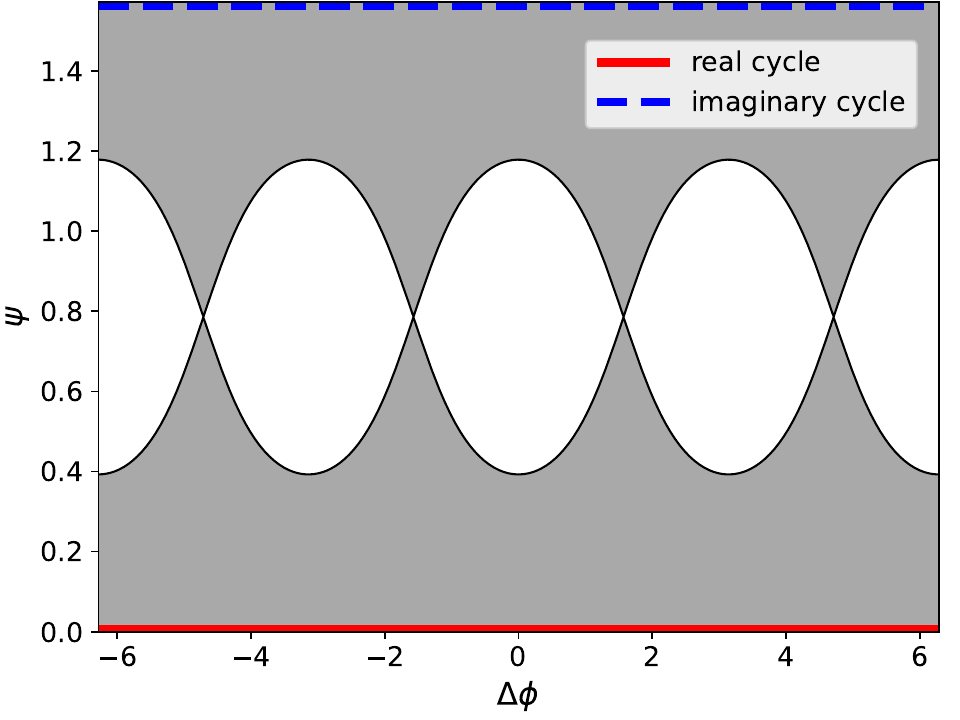}
	\caption{Sign of $P(\phi_x,\phi_y,\psi)$ in \eqref{eq:P} in the $(\Delta\phi,\psi)$ plane. In the shaded 
			 regions $P(\phi_x,\phi_y,\psi)>0$, such that $e^{-S}$ vanishes in the limit $r\to\infty$, while in 
			 the white regions $P(\phi_x,\phi_y,\psi)\leq0$. The real and imaginary integration cycles in these
			 coordinates are marked as full and dashed lines, respectively.}
	\label{fig:2D_cycles}
\end{figure}

Introducing the real and imaginary parts of $z_i$ as $z_i=x_i+\ii y_i$, we choose the following
-- convenient, but perhaps unusual -- parametrization:
\begin{align}\label{eq:parametrization}
	\begin{aligned}	
		x_1 &= r_x\cos(\phi_x)\;, \quad x_2 = r_x\sin(\phi_x)\;, \quad r_x = r\cos(\psi)\;, \\
		y_1 &= r_y\cos(\phi_y)\;, \quad y_2 = r_y\sin(\phi_y)\;, \quad  r_y = r\sin(\psi)\;,
	\end{aligned}
\end{align}
such that $\psi=\arctan\left(\frac{r_y}{r_x}\right)\in\left[0,\frac{\pi}{2}\right]$ .
Inserting \eqref{eq:parametrization} back into \eqref{eq:2d}, we obtain the
following expression for the real part of $S$, again assuming $\lambda_l=1$ without loss of generality:
\begin{equation}\label{eq:P}
	\mathrm{Re}\,S = r^4P(\phi_x,\phi_y,\psi)\;, \quad
	P(\phi_x,\phi_y,\psi) = \cos^2(2\psi)-\sin^2(2\psi)\cos^2(\phi_x-\phi_y)\;.
\end{equation}
Notice how the latter expression depends only on the difference $\Delta\phi:=\phi_x-\phi_y$, which nicely reflects
the $\mathrm{O}(2)$ symmetry of \eqref{eq:2d}. We have thus reduced the problem of finding the zeros of 
$e^{-S}$ to the problem of finding regions in $(\Delta\phi,\psi)$ space where $P(\phi_x,\phi_y,\psi)>0$. We show a 
plot of the sign of $P(\phi_x,\phi_y,\psi)$ in \cref{fig:2D_cycles}. 

The integration cycles can now be understood as incontractible paths within the $P>0$ regions, of which, as one can see 
from \cref{fig:2D_cycles}, there are only two. Taking into account the periodicity in 
$\Delta\phi$, this suggests that there exist (up to homology) $N_\gamma=2$ independent integration cycles, one 
for $\psi\lessgtr\frac{\pi}{4}$, respectively, and we have indicated these cycles in \cref{fig:2D_cycles}. Upon 
closer inspection, it turns out that the cycle in the region $\psi<\frac{\pi}{4}$ is equivalent to the real integration 
cycle $\gamma_1$, while the one for $\psi>\frac{\pi}{4}$ corresponds to the imaginary cycle $\gamma_2$. Here, we 
have defined (still assuming $\lambda_l=1$)
\begin{equation}
	\iint_{\gamma_1}dz_1dz_2 := \int_{-\infty}^\infty dz_1\int_{-\infty}^{\infty}dz_2\;, \quad
	\iint_{\gamma_2}dz_1dz_2 := \int_{-\ii\infty}^{\ii\infty} dz_1\int_{-\ii\infty}^{\ii\infty}dz_2\;.
\end{equation}
For general $\lambda_l$, the $\gamma_i$ are once again defined via appropriate rotations of the real and imaginary lines, as above.
It is curious to note that the number of independent integration cycles in the model \eqref{eq:2d} is, in fact, 
smaller than in its one-dimensional counterpart \eqref{eq:1d}. This observation, however, is due to the particular
definition of $S$ in \eqref{eq:2d}. For instance, the model $S=z_1^4+z_2^4$, which trivially decomposes into two 
independent one-dimensional models of the type \eqref{eq:1d}, features nine instead of two independent integration 
cycles. The fact that the amount of coupling between the degrees of freedom (drastically) affects the number of 
independent cycles as well as the special role of the $\mathrm{O}(2)$-symmetric model \eqref{eq:2d} 
will be discussed in detail in a forthcoming publication.

\begin{figure}[t]
	\centering
	\begin{subfigure}{.495\linewidth}
		\includegraphics[scale=0.45]{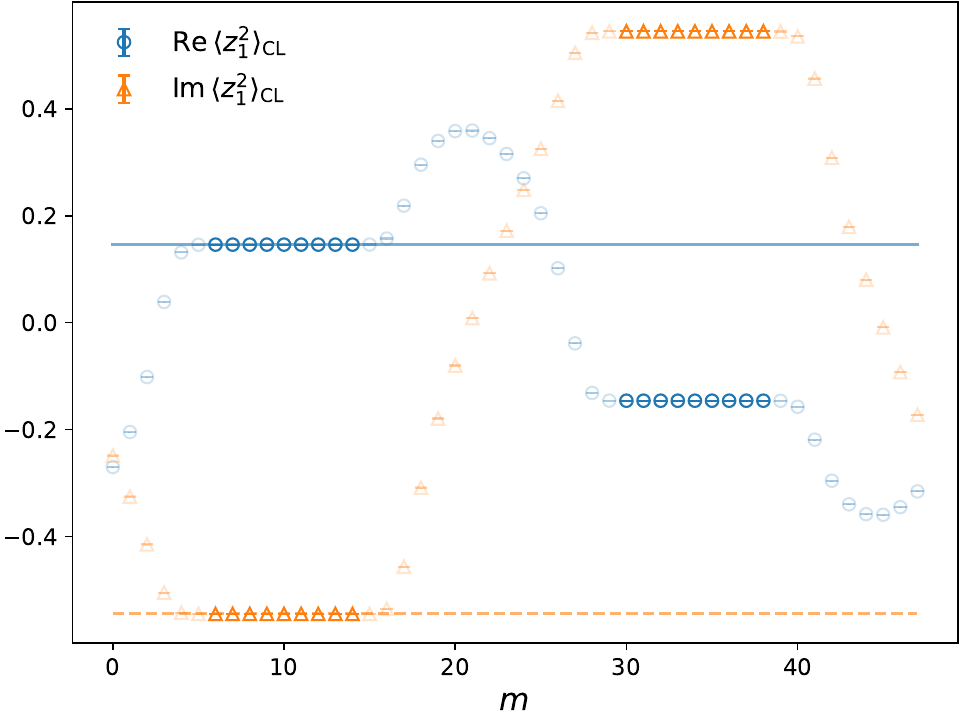}
	\end{subfigure}
	\begin{subfigure}{.495\linewidth}
		\includegraphics[scale=0.45]{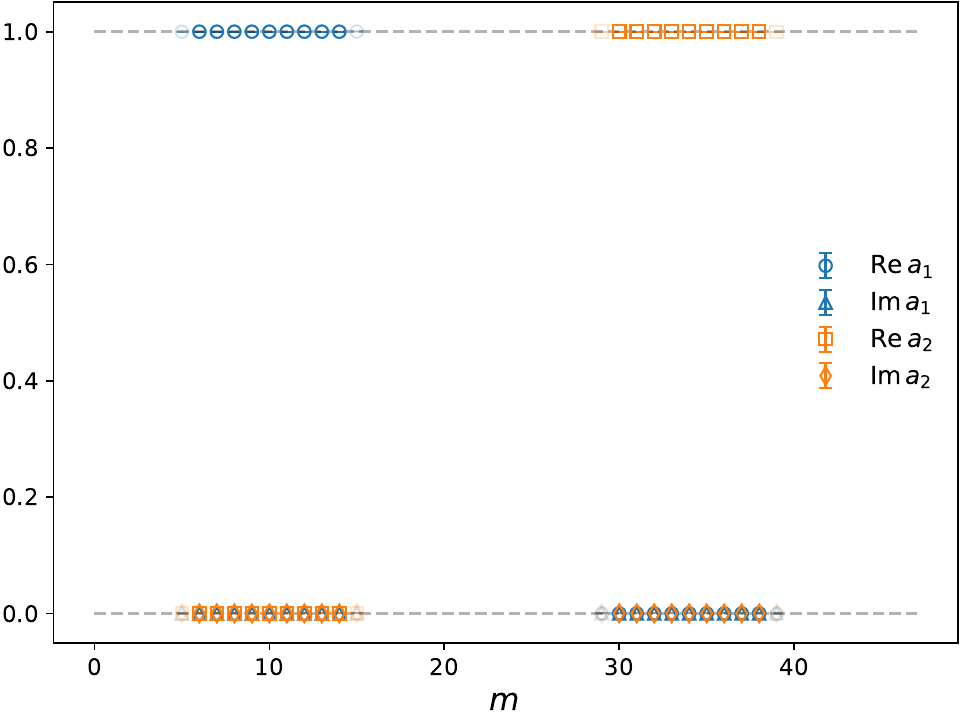}
	\end{subfigure}
	\caption{Analogous plots to \cref{fig:1D} for the two-dimensional model \eqref{eq:2d} with $l=5$. \emph{Left:}
			 Observable $\langle z_1^2\rangle$ as a function of $m$. \emph{Right:} The coefficients $a_i$ determined
			 from \eqref{eq:salcedo_seiler}, see the main text for details.}
	\label{fig:2D}
\end{figure}

Finally, let us discuss the validity of \eqref{eq:salcedo_seiler} for the two-dimensional case at
hand. We use a straightforward generalization of \eqref{eq:cle} to simulate the Langevin-time
evolution of $z_1$ and $z_2$, with a scalar kernel $K=K_m$ (see \eqref{eq:kernel}) that acts on $z_1$ and $z_2$ in the 
same way. We follow essentially the same steps as before, but now consider bi-variate monomial observables of the form
$\langle z_1^{n_1}z_2^{n_2}\rangle$. As an example analogous to \cref{fig:1D} (left), we show in \cref{fig:2D} 
(left) the dependence of $\langle z_1^2\rangle_\mathrm{CL}$ on $m$. There are striking similarities between the 
two figures as both show plateaus around $m=10$ and $m=34$, respectively. The other two plateaus present in 
\cref{fig:1D}, however, are not observed in the two-dimensional model. We thus conjecture that on the first and second 
plateau we should find $a_i=\delta_{i,1}$ and $a_i=\delta_{i,2}$, respectively, which also implies that both coefficients 
are entirely real and integer-valued. To substantiate this claim, we show in \cref{fig:2D} (right) the $m$-dependence of 
the coefficients $a_i$, determined from \eqref{eq:salcedo_seiler} using all observables 
$\langle z_1^{n_1}z_2^{n_2}\rangle_\mathrm{CL}$ for which $n_1+n_2\leq4$ and $n_i\neq3$. As expected, close to $m=10$ we 
indeed find $a_1=1$ and $a_2=0$ and \emph{vice versa} at $m=34$. We stress that -- as in the one-dimensional case -- we 
always obtain stable fits for the $a_i$ when the boundary terms vanish, as predicted by \eqref{eq:salcedo_seiler}.

The observations discussed here are first evidence that the theorem \eqref{eq:salcedo_seiler} can indeed be generalized
to higher dimensions. If this were the case, it could provide a valuable tool in advancing our knowledge on the
influence of kernels on complex Langevin simulations and -- in particular -- their convergence.

\acknowledgments
We are indebted to Enno Carstensen and Ion-Olimpiu Stamatescu for valuable discussions as well as past and ongoing 
collaborations. The numerical results presented in this work have been obtained in part in simulations on the 
computing cluster of the University of Graz (GSC) and the Vienna Scientific Cluster (VSC). This research 
was funded in part by the Austrian Science Fund (FWF) via the Principal Investigator Project 
\href{https://doi.org/10.55776/P36875}{P36875}. 
    
\bibliographystyle{JHEP}
\bibliography{bibliography}

\end{document}